\documentclass[%
 reprint,
 floatfix,
superscriptaddress,
 amsmath,amssymb,
 aps
]{revtex4-2}

\usepackage{graphicx}
\usepackage{dcolumn}
\usepackage{bm}
\usepackage{subfig}

\begin{document}

\preprint{APS/123-QED}

\title{Deceleration and trapping of SrF molecules}
 
\author{P.~Aggarwal}
\author{Y.~Yin}
\author{K.~Esajas}
\affiliation{Van Swinderen Institute for Particle Physics and Gravity, University of Groningen, The Netherlands}
\affiliation{Nikhef, National Institute for Subatomic Physics, Amsterdam, The Netherlands}

\author{H.L.~Bethlem}
\affiliation{Van Swinderen Institute for Particle Physics and Gravity, University of Groningen, The Netherlands}
\affiliation{Department of Physics and Astronomy, and LaserLaB, Vrije Universiteit Amsterdam, The Netherlands}
\author{A.~Boeschoten}
\author{A.~Borschevsky}

\affiliation{Van Swinderen Institute for Particle Physics and Gravity, University of Groningen, The Netherlands}
\affiliation{Nikhef, National Institute for Subatomic Physics, Amsterdam, The Netherlands}
\author{S.~Hoekstra}
 \email{Electronic mail: s.hoekstra@rug.nl}
\affiliation{Van Swinderen Institute for Particle Physics and Gravity, University of Groningen, The Netherlands}
\affiliation{Nikhef, National Institute for Subatomic Physics, Amsterdam, The Netherlands}
\author{K.~Jungmann}
\author{V.R.~Marshall}
\author{T.B.~Meijknecht}
\affiliation{Van Swinderen Institute for Particle Physics and Gravity, University of Groningen, The Netherlands}
\affiliation{Nikhef, National Institute for Subatomic Physics, Amsterdam, The Netherlands}
\author{M.C.~Mooij}
\affiliation{Nikhef, National Institute for Subatomic Physics, Amsterdam, The Netherlands}
\affiliation{Department of Physics and Astronomy, and LaserLaB, Vrije Universiteit Amsterdam, The Netherlands}
\author{R.G.E.~Timmermans}
\author{A.~Touwen}
\affiliation{Van Swinderen Institute for Particle Physics and Gravity, University of Groningen, The Netherlands}
\affiliation{Nikhef, National Institute for Subatomic Physics, Amsterdam, The Netherlands}
\author{W.~Ubachs}
\affiliation{Department of Physics and Astronomy, and LaserLaB, Vrije Universiteit Amsterdam, The Netherlands}
\author{L.~Willmann}
\affiliation{Van Swinderen Institute for Particle Physics and Gravity, University of Groningen, The Netherlands}
\affiliation{Nikhef, National Institute for Subatomic Physics, Amsterdam, The Netherlands}

\collaboration{NL-$e$EDM Collaboration}

\date{\today}

\begin{abstract}
We report on the electrostatic trapping of neutral SrF molecules. The molecules are captured from a cryogenic buffer-gas beam source into the moving traps of a 4.5~m long traveling-wave Stark decelerator. The SrF molecules in $X^2\Sigma^+(v=0, N=1)$ state are brought to rest as the velocity of the moving traps is gradually reduced from 190 m/s to zero. The molecules are held for up to 50 ms in multiple electric traps of the decelerator. The trapped packets have a volume (FWHM) of 1~mm$^{3}$ and a velocity spread of 5(1)~m/s which corresponds to a temperature of $60(20)$~mK. Our result demonstrates a factor 3 increase in the molecular mass that has been Stark-decelerated and trapped. Heavy molecules (mass$>$100~amu)  offer a highly increased sensitivity to probe physics beyond the Standard Model. This work significantly extends the species of neutral molecules of which slow beams can be created for collision studies, precision measurement and trapping experiments.
\end{abstract}

\maketitle

Slow beams and trapped samples of heavy diatomic molecules are highly interesting as probes of fundamental physics, specifically for a measurement of the electron's electric dipole moment (eEDM) \cite{Hudson2011, Cairncross2017, Vutha2018, Andreev2018,  Aggarwal2018} and to study parity violation \cite{Altunta2018}. However, it is challenging to obtain slow beams or even trapped samples of suitable molecules, which are key to achieve increased precision in spectroscopy \cite{Tarbutt2009}.

Stark deceleration is a successful technique to capture molecules from a supersonic expansion and decelerate them \cite{Meerakker2008}. Beams of ND$_{3}$ \cite{Bethlem2000nature},  NH$_{3}$ \cite{Bethlem2002}, OH \cite{Meerakker.Meijer.2005mda, Sawyer.Ye.2007}, OD \cite{Hoekstra2007}, NH \cite{Hoekstra.Meerakker.20074}, CO~\cite{Gilijamse2007}, CH$_{3}$F \cite{Congsen2015} molecules have been decelerated and trapped using this technique. Some of these decelerated and trapped samples have been used for high-resolution spectroscopy \cite{Veldhoven2004}, the measurement of collision cross-sections \cite{Gilijamse2006} and the determination of lifetimes of long-lived electronically and vibrationally excited states \cite{Meerakker2005,Gilijamse2007}. Magnetic \cite{Vanhaecke2007, Narevicius2008} and optical \cite{Fulton2006} analogues of the decelerator have also been developed. Until now, CH$_{3}$F \cite{Congsen2015} and O$_{2}$  \cite{Akerman2017} are the heaviest molecules to have been trapped following Stark and Zeeman deceleration, respectively. The heavy alkaline-earth monohalide molecules, like SrF and BaF (with a mass of 107 and 156 amu, respectively), are especially interesting for the precision measurements mentioned above \cite{Altunta2018, Aggarwal2018}. However, their energy-level structure and the resulting Stark shifts limit the possible deceleration strength that can be applied in the decelerator. This, together with the larger mass, requires a much longer decelerator - but especially at low velocities, switching-type decelerators suffer from instabilities \cite{Meerakker2006} which stand in the way of high-intensity slow beams and trapped samples.

We combine two developments to make the deceleration and trapping of these molecules possible. Firstly, 2nd generation decelerators, such as the traveling-wave Stark decelerator (TWSD) \cite{Osterwalder2010, Meek.Osterwalder.2011, Berg.Meinema.2014, Mathavan2016, Quintero-Perez.Bethlem.2013c9q} and advanced operation of the switching-type decelerators~\cite{Scharfenberg2009, Reens.Ye.2020}, allow the stable operation of long decelerators. For this purpose we have built a 4.5 meter long TWSD. Secondly, cryogenic buffer-gas molecular beam sources have been developed that, compared to supersonic beams, produce molecules with comparatively higher intensities and a significantly lower velocity of 150-200 m/s \cite{Hutzler2012}. The combination of the intense, slow beam provided by the source and the stability of TWSD has allowed us to decelerate SrF to standstill.

The alkaline-earth monohalides also have been found to  be amenable to laser cooling \cite{Fitch2021}. Especially the CaF molecule has a highly closed system of energy levels, enabling the repeated scattering of many photons. The heavier molecules SrF \cite{Barry2012}, BaF \cite{Hao2019} and also RaF \cite{Isaev2010} are suitable but with increasing leaks out of the closed cycle, requires more repump lasers to close the leaks to other hyperfine, rotational, vibrational and even electronic levels. To date, laser cooling and magneto-optical traps have been demonstrated for SrF \cite{Barry2012, Barry2014}, CaF \cite{Zhelyazkova2014, Hemmerling2016, Anderegg2017, Williams2017} and YO molecules \cite{Hummon.Ye.2013, Collopy2018}, which have been loaded from cryogenic buffer-gas sources. The transverse lasercooling of YbF molecule, which is used for probing eEDM \cite{Hudson2011, Ho2020}, has also been demonstrated \cite{Lim2018}. For the BaF molecule, the combination of Stark deceleration with transverse laser cooling (requiring only a limited number of photons to be scattered) is a promising approach to create an intense, slow and cold beam for a highly sensitive eEDM measurement~\cite{Aggarwal2018}. 

\begin{figure*}[t]
	\subfloat{\includegraphics[width=0.6\textwidth]{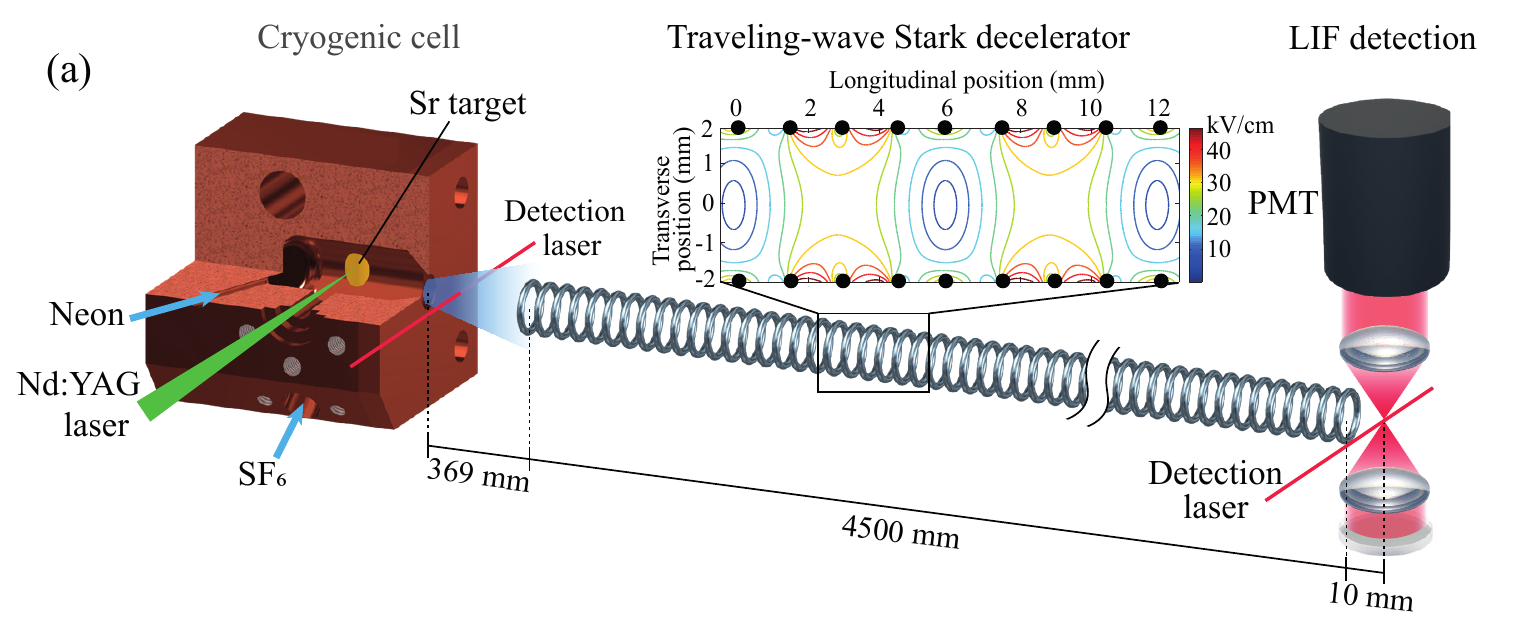}}
	\hspace{1em}%
	\subfloat{\includegraphics[width=0.34\textwidth]{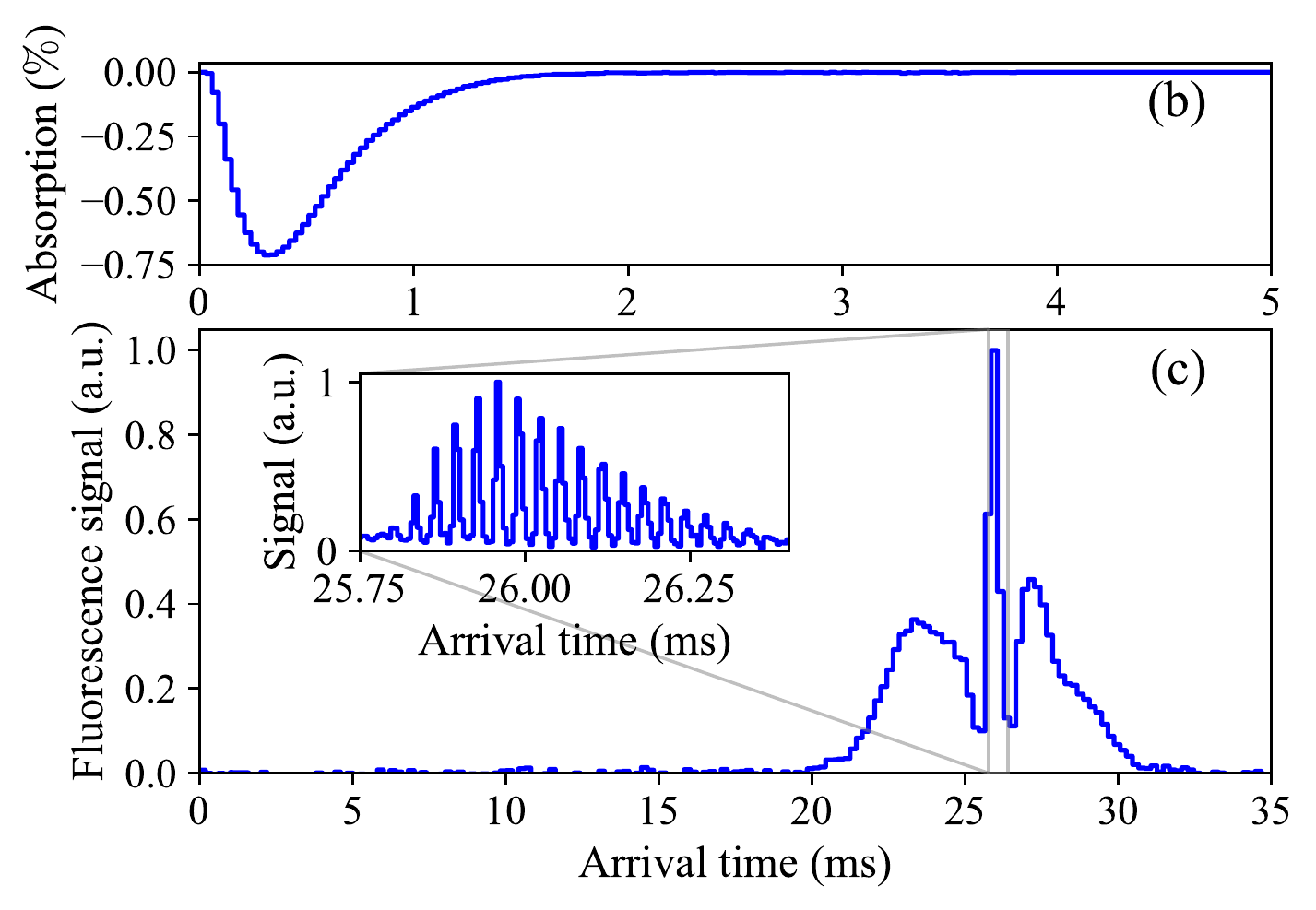}}
	\caption{\label{fig:setup} (a) Experimental setup: The SrF molecules are produced in a cryogenic buffer gas source. A high energy pulsed Nd: YAG laser ablates a strontium metal target. The ablation products of the strontium metal react with the SF$_{6}$ gas to produce SrF, which is then cooled on collisions with the cold neon gas inside the cell. On exiting from the cell, molecules enter the 4.5-m-long traveling-wave Stark decelerator (TWSD) that is mounted 36.9 cm behind the source exit. The molecules travel through the decelerator in the guided or the deceleration mode. They are then detected at a distance of 1 cm from the last ring electrode of the TWSD. (b) The typical absorption signal of the molecular beam measured at a distance of 5~mm from the cell exit. (c) The experimental time-of-flight profile with a binsize of 200~$\mu$s, shows the guiding of the molecular beam of SrF with an average velocity of 190 m/s at 5.0 kV. Inset: a zoom-in of the guided peak with a binsize of 5~$\mu$s, which shows the individual electric field traps filled by the molecules inside the TWSD.}
\end{figure*}

In this paper, we demonstrate the deceleration and trapping of SrF molecular beam from a cryogenic source in a 4.5 m long TWSD. The experimental setup, which is shown in Fig.~\ref{fig:setup}(a), consists mainly of three parts: a cryogenic buffer gas source, a TWSD, and a laser-induced-fluorescence (LIF) detection section. SrF molecules are produced and cooled in the cryogenic buffer gas source that uses a two-stage pulse tube cryocooler. The first stage is held at a temperature of 24-26 K and the second stage at a temperature of 4 K. A copper cell whose design is adopted from \cite{Truppe2017} is attached to the second stage of the cryocooler, and the temperature of the cell is maintained at 17~K via a heater. A 4~mJ, 5~ns laser pulse with a repetition rate of 10~Hz from a Nd:YAG laser at 532~nm ablates a strontium metal target mounted inside the cell. Sulphur hexafluoride (SF$_{6}$) gas enters the cell with a flow rate of 0.5 sccm, where it reacts with the ablation products to form SrF. The SrF molecules thermalise with pre-cooled neon which enters the cell from a different gas line at a flow rate of 8.0 sccm. Molecules are detected at a distance of 5 mm from the cell exit via absorption on the 663 nm transition $A^2\Pi_{1/2}(v'=0, J'=1/2)$ $\leftarrow$ X$^2\Sigma^+(v=0, N=1)$. The hyperfine structure of the $N =1$ rotational level in the ground electronic state $ X^2\Sigma^+(v=0, N=1)$ is covered by adding sidebands to the detection laser. A typical absorption signal of the molecules is shown in Figure~\ref{fig:setup}(b).

As shown in the inset of Figure~\ref{fig:setup}(a), in the 4.5 m long TWSD, a time-dependent inhomogeneous electric field distribution is created inside 3024 ring-shaped electrodes with a diameter of 4 mm, as previously described in ~\cite{Mathavan2016}. A sinusoidal voltage is applied individually to every eight electrodes: $V_{n}(t) = V_{0}\sin(-\phi(t) + 2\pi n/8)$ where $n = 0,1,2....7$. The decelerator can be operated in two different modes, the guiding mode and the deceleration mode depending on the phase $\phi(t)$. During the guiding mode, the phase is $\phi(t) = 2\pi ft$ where $f$ is a constant frequency of the sinusoidal sine wave which translates into a constant velocity of the moving electric field minima along the axis of the decelerator. In the deceleration mode, the frequency of the sine-wave is chirped down, thereby reducing the speed of the electric field minima. The velocities of SrF molecules in the $X^2\Sigma^{+}(v=0, N =1)$ low-field seeking state can thus be manipulated accordingly.

Molecules from the Stark decelerator are detected using laser-induced fluorescence (LIF).  About 1~cm away from the last ring of the decelerator, a diode laser beam with a diameter of 0.7~mm and a power of 0.5~mW interacts perpendicularly with the molecular beam, exciting the molecules on the same transition as is used for the absorption detection. The fluorescence emitted from molecules is collected and imaged onto a PMT by a combination of lenses with a collection solid angle of 0.93~steradian and a collection efficiency of 7.5$\%$. The molecular time-of-flight signal is then recorded by making a histogram of the fluorescence photon counts with respect to the arrival time of the molecules, with $t=0$~ms corresponding to the time at which the Nd:YAG laser ablates the strontium target.
 
We determined the central velocity of the molecular beam from the cryogenic source by guiding the molecules at different speeds and comparing the strengths of the LIF signals. The guiding velocity which yields the highest intensity in the guiding signal peak was determined to be 190 m/s. This velocity is also employed later as the initial velocity for the deceleration mode of the TWSD.

Figure~\ref{fig:setup}(c) shows a typical time-of-flight LIF signal of the molecules guided at 190~m/s with a 5~kV amplitude sinusoidal voltage applied to the decelerator. A central peak arrives 26~ms after ablation, corresponding to the molecules which are within the longitudinal and transverse phase-space acceptance of the decelerator. These molecules can thus be trapped and guided till the end. The two adjacent wings correspond to the molecules that are outside the longitudinal but inside the transverse acceptance. These molecules are confined transversely while traveling through the decelerator.

A zoom-in of the guided peak, as shown in the inset of Figure~\ref{fig:setup}(c), indicates that there are approximately sixteen sub-peaks within it, which correspond to the molecules trapped in sixteen consecutive electric wells inside the decelerator. This is a consequence of the relatively long pulse of molecules from the cryogenic buffer-gas source, in contrast to the guiding of a supersonic molecular beam in the same decelerator \cite{Mathavan2016}. As the periodicity of the electric wells is 6~mm (see Fig.~\ref{fig:setup}(a)), the guided molecular beam packet extends to $\sim 100$~mm.

\begin{figure}
\includegraphics[scale=0.5]{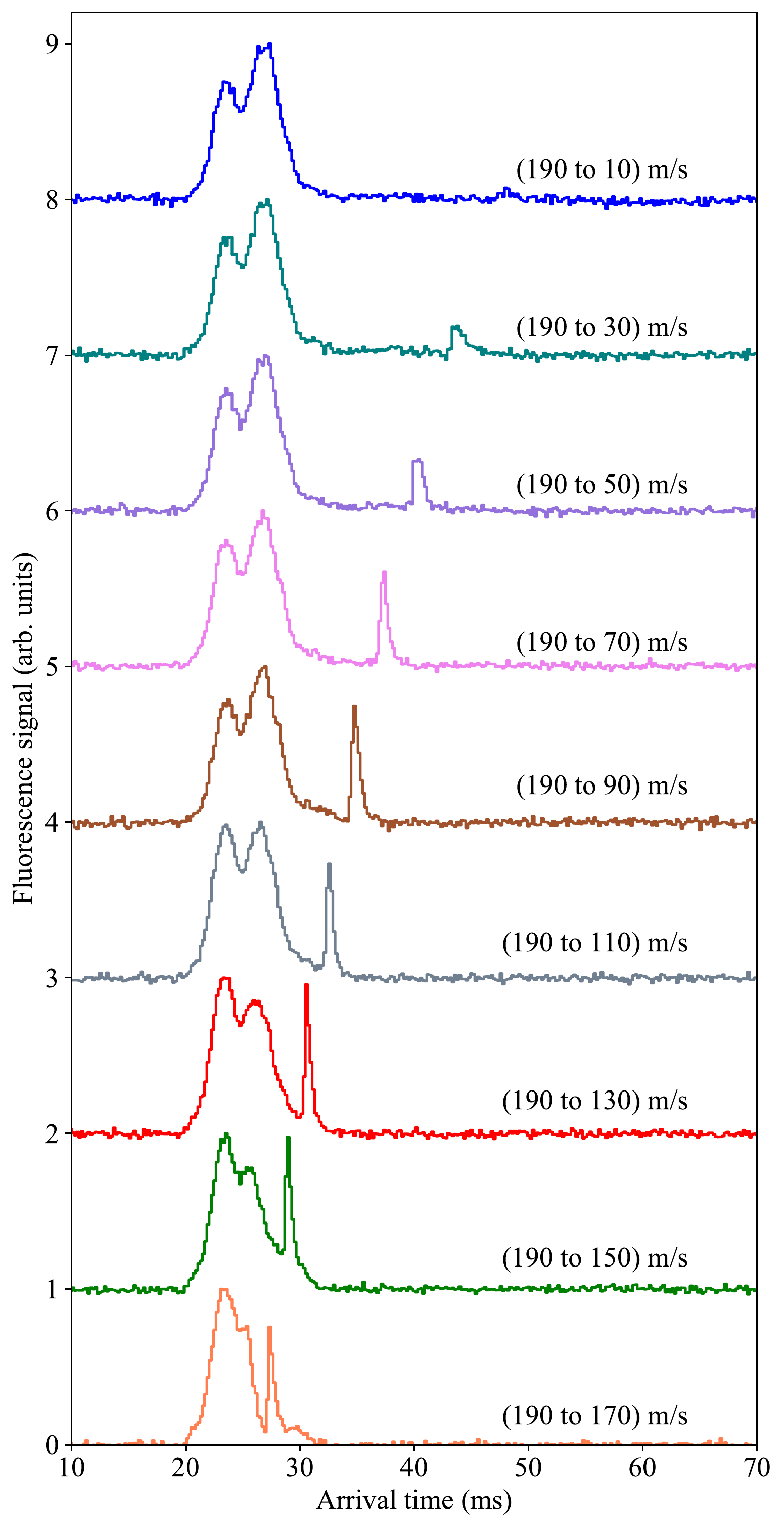}
\caption{\label{fig:decelerate_exp}Time-of-flight LIF signals depicting the deceleration of SrF molecules from 190 m/s to different final velocities till 10 m/s at a voltage amplitude applied to the decelerator of 5 kV with a binsize of 200~$\mu$s . A vertical offset has been added to different plots for clarity.}
\end{figure}

We measured the deceleration of the molecular beam from 190 m/s to a series of final velocities ranging from 170 to 10 m/s. The corresponding time-of-flight signals are shown in Figure~\ref{fig:decelerate_exp} with a binsize of 200~$\mu$s, where each plot is averaged over 12000 ablation shots and a vertical offset has been added to the plots for clarity. There are molecules arriving between 20~ms and 30~ms, which corresponds to the ones that failed to be trapped longitudinally but were confined transversely (we will refer to them as non-slowed part of the signal). A fraction of molecules arrive at a later time as we decrease the final velocity of deceleration (which will be referred to as slowed part). The height of the non-slowed part remains basically the same for all the plots, although the distributions of them are slightly different due to the small variation of the molecular beam properties over our measurement time. The height of the slowed part decreases with a lower final velocity. This is due to the decrease in longitudinal acceptance of the decelerator with an increasing deceleration strength and an increase in divergence of the beam at lower velocities. The width of the slowed part increases for deceleration to lower final velocities because the free flight time between the last ring of decelerator and LIF detection point increases. 

\begin{figure}
\includegraphics[scale=0.6]{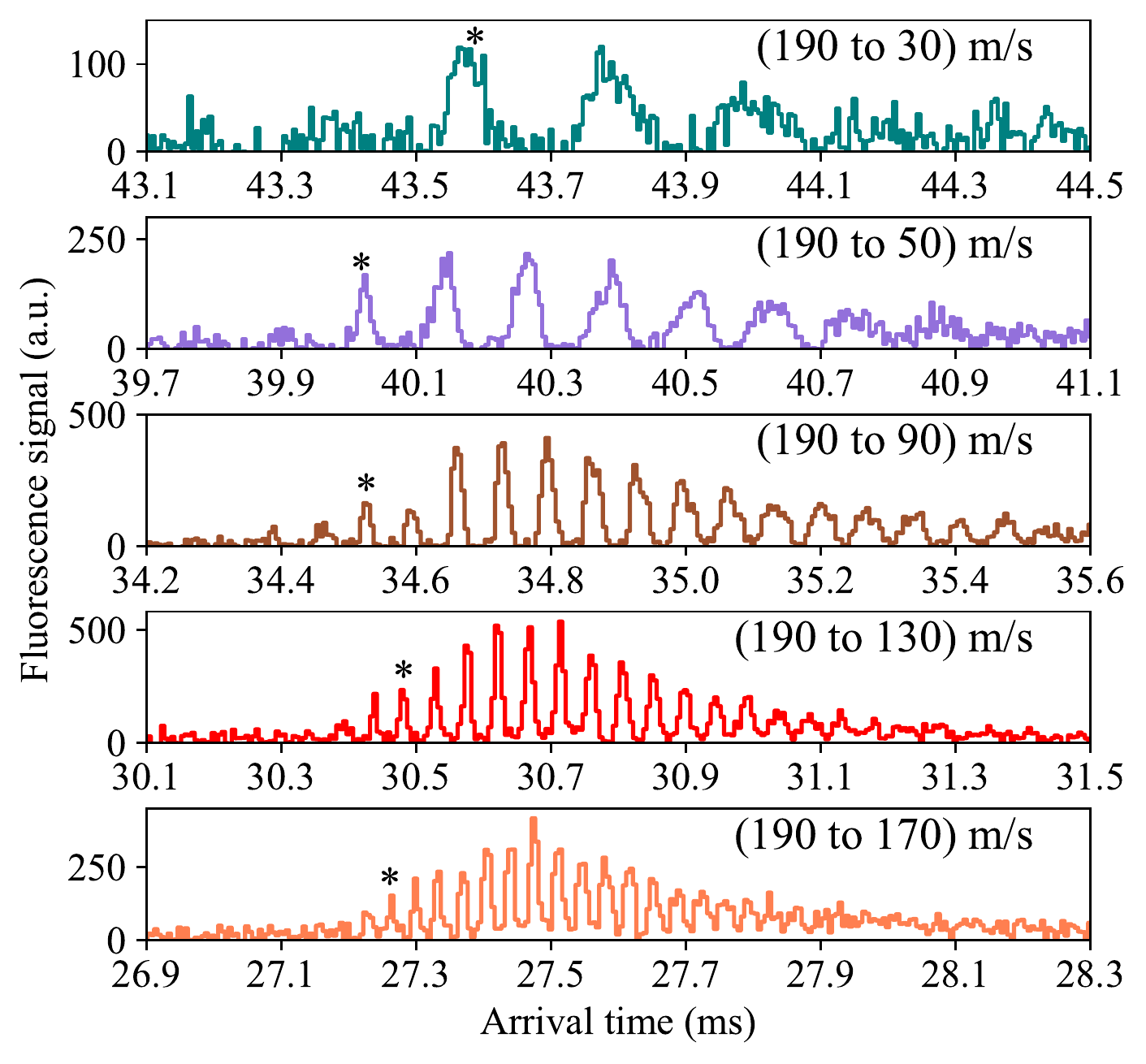}%
\caption{\label{fig:decelerate_exp_zoom}Zoom in of the time-of-flight histograms around the arrival time  of the decelerated peaks with a binsize of 5~$\mu$s, for a few selected final velocities 170, 130, 90, 50, 30 m/s. The peaks labeled (*) correspond to the synchronous molecules, which represent the first packet of molecules that is detected after the electric fields of the decelerator are switched off.}
\end{figure}

To investigate the velocity spread of the decelerated molecules, a zoom-in of the slowed part for a number of different final velocities is shown in Fig.~\ref{fig:decelerate_exp_zoom}. The peaks correspond to molecules in multiple traps. For slower molecules these peaks are broader, as a result of the longer flight time from the moment the decelerator is switched off to the detection point. For the slower beams, we can identify fewer molecular packets, which we attribute to their increased divergence. The width of the adjacent peaks for a certain final velocity increases with their arrival time, which is due to the longitudinal expansion of the molecular packet after the decelerator is switched off~\cite{Veldhoven2006}. The labeled peak (*) in Fig.~\ref{fig:decelerate_exp_zoom} corresponds to the synchronous molecules, which represent the first molecular packet that is detected after the electric fields of the decelerator are switched off. By calculating the width difference for adjacent peaks and the difference in their arrival time, the FWHM (full width at half maximum) velocity spread of the slowed molecules is 10(2)~m/s for molecules decelerated to 170~m/s and 4(1)~m/s for molecules decelerated to 50~m/s. The corresponding position spreads of the beams decelerated to 170 m/s and 50 m/s are 2.1(3)~mm and 1.0(2), respectively. The uncertainty of these numbers takes into account the ramp down time to switch off the electric field on the decelerator (260 $\mu s$).

\begin{figure}
\includegraphics[scale=0.5]{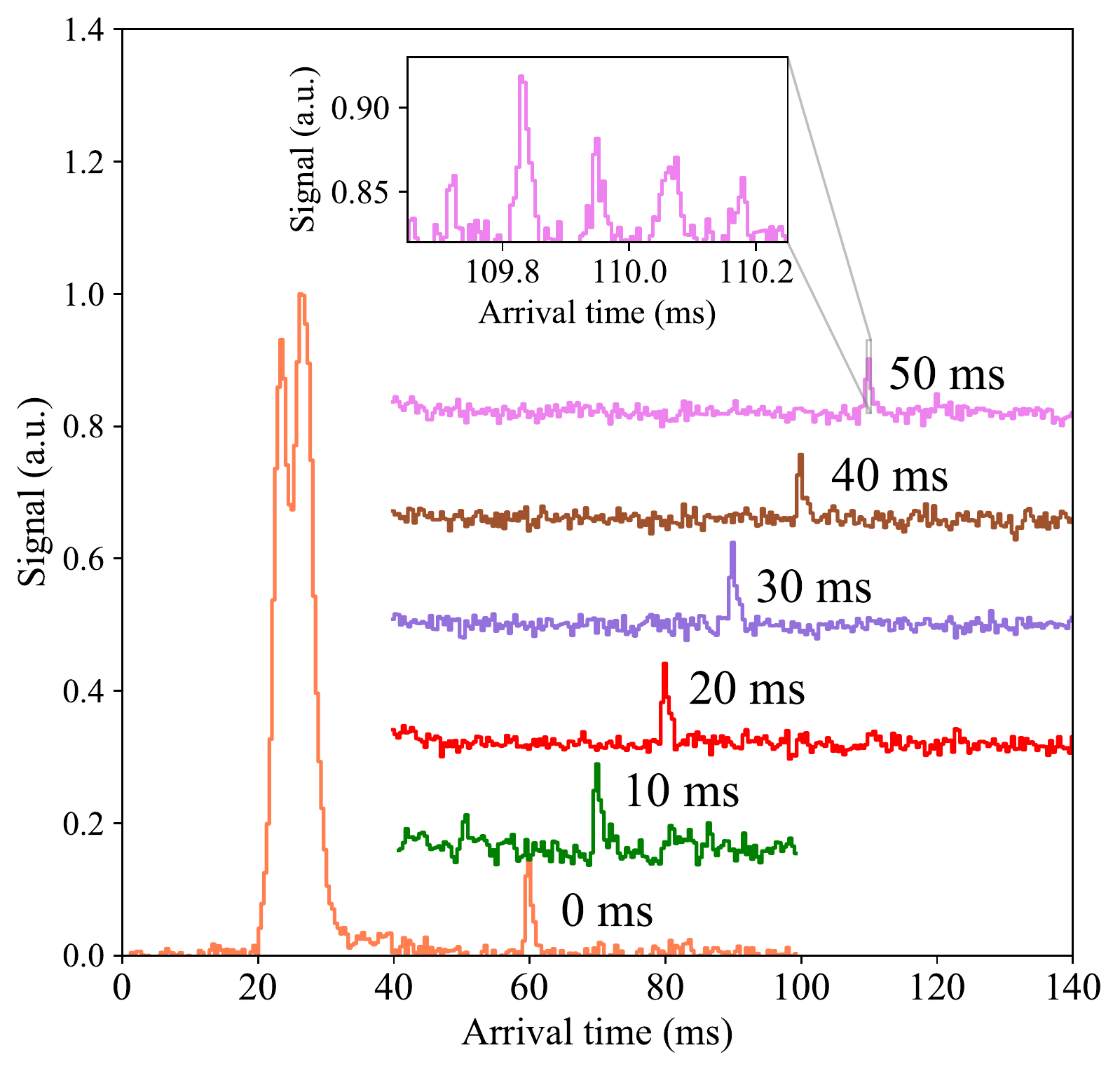}%
\caption{\label{fig:trapping_data} Time-of-flight histograms  with a binsize of 400~$\mu$s demonstrating the trapping of SrF molecules for different trapping times $\Delta t$. Molecules are decelerated from 190 m/s to 0 m/s in 4.2m, and then held in the electric traps of the TWSD for $\Delta t$ = 0, 10, 20, 30, 40 and 50 ms, and then accelerated to 52 m/s in the remaining 0.3~m of the TWSD. The undecelerated molecules are only shown for the case of $\Delta t = 0$. A vertical offset has been added to the plots for clarity. The inset shows a zoom-in of the trapping peak for the case of $\Delta t = 50$ ms.}
\end{figure}

In Fig.~\ref{fig:trapping_data}, we demonstrate the stopping and trapping of the SrF molecules in the lab frame inside the electric traps of the decelerator. SrF molecules are decelerated from 190~m/s to 0~m/s over a length of 4.2~m and held for a trapping time $\Delta t$ in the inherent electric traps of the decelerator. To detect the trapped molecules they are accelerated out of the decelerator with the same magnitude of acceleration strength as the deceleration strength, meaning that molecules are accelerated to 52 m/s in the remaining 0.3~m length of the decelerator and detected 1~cm away from the last ring of the decelerator. The acceleration of SrF molecules trapped for different trapping times to the same final velocity also facilitate good quantitative comparison of the fluorescence signal strength.

The time-of-flight profile for different trapping times ($\Delta t$) with a binsize of 400 $\mu$s is shown in Fig~\ref{fig:trapping_data}. Each plot has been averaged over 72000 shots and a vertical offset is added to subsequent time-of-flight profiles for clarity. The molecules between the arrival time of 20-30~ms, similar to the case of deceleration, are those confined in the decelerator only transversely but not longitudinally. A signal peak at 60~ms for the first case demonstrates a case of zero trapping time ($\Delta t$=0~ms), where the molecules are decelerated to zero and immediately accelerated out. The subsequent plots show the molecules trapped for different trapping time between 10-50~ms, resulting in molecules arriving later. A zoom-in of the time-of-flight profile around the arrival time of the molecules trapped for 50~ms with a binsize of 5~$\mu$s is shown in the inset of Fig~\ref{fig:trapping_data}. With the current signal-to-noise, we clearly observe molecules trapped in at least 5 neighbouring electric field traps inside the decelerator but from the measurements presented in Fig.~\ref{fig:decelerate_exp_zoom}, we know that $\sim~16$ traps should be filled. The axial and the radial width of a single trap is 1.0(2)~mm. The FWHM velocity spread of the trapped molecules is deduced to be about 5(1)~m/s, which corresponds to a translational temperature of 60(20)~mK. 
 
To characterise the performance of the TWSD, we determine the fraction of photon signal detected from the decelerated molecules. This fraction varies between 13$\%$ for deceleration to 170~m/s and 1.2$\%$ for deceleration to 10~m/s. This fraction reflects both the increased divergence of the beam and the reduced phase-space acceptance of the TWSD~\cite{Berg.Meinema.2014} for deceleration to lower velocities. For the trapping data, $(2.8-1.3)\%$ are trapped for $0-50$~ms inside the electric traps of the decelerator.

The average number of total detected photons per shot ranges from $10-30$ in the datasets presented in this paper. We estimate the number of molecules in the beam per detected photon to be $10^2-10^3$, based on an estimate of the detection volume, fluorescence collection efficiency, divergence of the beam and excitation efficiency taking into account the stray electric field in the detection region. When combining these numbers with the fraction of photons from decelerated and trapped molecules, we find that the number of molecules in the trapped samples and the slow beam ranges between $10-4000$. We can also estimate the number of decelerated molecules using the density of the molecular beam measurements combined with the calculated phase space acceptance of the decelerator. From the absorption signal shown in Fig.~\ref{fig:setup}(b), we estimate that $\sim 10^{9}$ SrF molecules are produced in the $N=1$ state. Of these molecules, a fraction of $1.5\times10^{-6}$ have a position and velocity that falls within the acceptance of the decelerator, hence, we expect to decelerate on the order of 1000 molecules per shot, in reasonable agreement with the numbers derived from the LIF data.

Potential ways to increase the molecular numbers include implementing a better matching of the cryogenic source to the decelerator with a hexapole guide, by optically pumping the molecules into the state of interest before they enter the decelerator~\cite{Ho2020}, and by increasing the electric field strength of the decelerator to allow deceleration in the $(N=2)$ state.

In conclusion, we demonstrated the deceleration and trapping of a SrF molecular beam in the state $X^2\Sigma^+(v=0, N=1)$  produced in a cryogenic buffer gas source in a 4.5 m-long TWSD. SrF molecules with an average velocity of 190 m/s are decelerated down to arbitrarily low velocities, and are brought to a standstill and trapped for about 50 ms in a series of consecutive electric wells inside the decelerator. 
Although the SrF molecule is also amenable to laser cooling and trapping, the technique we employ here relies only on its Stark shift. Therefore, this method gives access to molecular species for which laser cooling is not feasible, opening the prospects for controlling more diverse and complex molecular species desirable for many applications in quantum science and technology and new physics searches. Also lighter molecules with an unfavourable Stark shift over mass ratio~\cite{Motsch.Pinkse.2009} can now be decelerated and trapped. The sensitivity to the eEDM, that scales with the third power of the atomic number, furthermore increases linearly with the long interaction time offered by cold and slow beams of heavy molecules. We will therefore apply the principle of the method demonstrated here to BaF, which is a prime candidate to probe physics beyond the Standard Model of particle physics~\cite{Aggarwal2018}.

P. Aggarwal and Y. Yin have contributed equally to this work. The NL-$e$EDM consortium receives program funding (EEDM-166) from the Netherlands Organisation for Scientific Research (NWO). We acknowledge support from Mike Tarbutt and Stefan Truppe in the design and construction of the cryogenic source. We thank Leo Huisman for technical assistance to the experiment.

\nocite{*}


%

\end{document}